\documentclass[aps,prd,superscriptaddress,floatfix,nofootinbib,showkeys,showpacs,twocolumn,reprint]{revtex4-1}

\usepackage{graphicx}
\usepackage{cancel}
\usepackage{amssymb}
\usepackage{textcomp}
\usepackage{amsmath}
\usepackage{bm}
\usepackage{times}
\usepackage{epsfig}
\usepackage{epstopdf}
\usepackage{color}
\usepackage{multirow}
\usepackage[colorlinks=true, pdfstartview=FitV, linkcolor=blue, citecolor=blue, urlcolor=blue]{hyperref}

\def\lsim{\mathrel{\raise.3ex\hbox{$<$\kern-.75em\lower1ex\hbox{$\sim$}}}}
\def\gsim{\mathrel{\raise.3ex\hbox{$>$\kern-.75em\lower1ex\hbox{$\sim$}}}}

\definecolor{orange}{rgb}{1,0.5,0}

\newcommand{\be}{\begin{equation}}
\newcommand{\ee}{\end{equation}}
\newcommand{\bea}{\begin{eqnarray}}
\newcommand{\eea}{\end{eqnarray}}

\newcommand{\minigraph}[5][0.25in]{\begin{minipage}{#2}\begin{center}\includegraphics[width=#2]{#5}\\\vspace{#3}\hspace{#1}{\footnotesize #4}\end{center}\end{minipage}}

\begin{document}

\title{Constraints on dimension-seven operators with a derivative in effective field theory for Dirac dark matter}

\author{Tong Li}
\email{litong@nankai.edu.cn}
\affiliation{
School of Physics, Nankai University, Tianjin 300071, China
}
\author{Yi Liao}
\email{liaoy@nankai.edu.cn}
\affiliation{
School of Physics, Nankai University, Tianjin 300071, China
}
\affiliation{
Center for High Energy Physics, Peking University, Beijing 100871, China
}

\begin{abstract}
The effective field theory (EFT) for dark matter (DM) has been widely used to investigate dark matter detection in both theoretical prediction and experimental analysis. To form a complete basis of effective operators for Dirac DM EFT at dimension seven, eight new four-fermion operators with a derivative in DM currents have recently been introduced. We discuss the experimental observables and constraints for the theoretical predictions of these new operators to constrain the DM mass and relevant energy scale. The observables from thermal relic abundance, indirect and direct detection, and LHC constraints are presented.
\end{abstract}


\maketitle

\section{Introduction}
\label{sec:Intro}

Despite of substantial efforts in theory and experiment, the microscopic properties of dark matter (DM) particles are still unknown. Due to the plethora of competing theoretical models in the current status, it is not feasible to extract the fundamental properties of dark matter particles by contrasting theoretical predictions with observation. To avoid this problem, the effective field theory (EFT) offers a good approach based on general and minimal theoretical assumptions regarding the physics underlying dark matter particles~\cite{Beltran:2008xg,Fan:2010gt,Goodman:2010qn,Balazs:2014rsa}. The EFT principle formulates specific dark matter models as a quantum field theory, under the assumption that the DM candidate is a single particle beyond the Standard Model (SM) and all other degrees of freedom are either heavy enough to be integrated out or have negligible strength for the observable spectrum.

The simplest way to build an EFT for DM is to introduce a new SM gauge singlet field, $\chi$. We assume it to be a Dirac fermion in this article. We assume further it is odd under a new parity while all SM fields are even, so that the $\chi$ particle is guaranteed to be stable and can only be created or annihilated in pair. The interaction Lagrangian containing all Lorentz and gauge invariant operators involving a pair of the $\chi$ field is schematically written as an expansion in the dimension $d$ of operators
\begin{eqnarray}
\mathcal{L}_\chi = \sum_{d,i,f}\mathcal{C}_{i}^d\mathcal{O}_{i,f}^d \ ,
\end{eqnarray}
where the Wilson coefficients are generically parameterized as $\mathcal{C}_{i}^d=\Lambda_{i}^{4-d}$ with $\Lambda_i$ being an effective cutoff scale for EFT. We will restrict ourselves in this work to the interactions of DM with the quarks, and assume the coefficients are universal in quark flavor $f=q$. The framework includes two electromagnetic dipole operators at dimension-5 (dim-5) and four dim-6 operators formed from products of vector and axial-vector currents~\cite{Goodman:2010qn}. Those dim-6 operators have been adopted to analyze the data of DM searches in indirect (ID) and direct detection (DD) experiments. For the unsuppressed vector operators, the DD measurements result in more stringent limits than what can be expected from relic abundance~\cite{Balazs:2017hxh}. Moreover, the spin-1 mediator scenario via a vector or axial-vector interaction is highly constrained by the $Z'\to$ dijet searches at the Large Hadron Collider (LHC), and the mass of the (axial-)vector mediators has been excluded up to a scale of 4-5~TeV~\cite{Khachatryan:2016ecr,Sirunyan:2016iap,
Sirunyan:2017nvi,Sirunyan:2018wcm,Aaboud:2017yvp,
Aaboud:2018zba}. The largely pushed energy scale for dim-6 operators motivates us to consider the effects of higher dimensional operators. The higher dimensional operators could be induced by underlying theories at perhaps a different energy scale from dim-6 operators, and thus may likely dominate the DM relic abundance when only one operator is switched on at a time. In fact, the detailed phenomenology of the six dim-7 operators involving a scalar or tensor fermion current has been widely investigated~\cite{DAmbrosio:2002vsn,Goodman:2010ku,
Goodman:2010qn}. For instance, for the scalar operator scaled by quark mass, the DD experiments and the LHC have excluded the energy scale below about 1 TeV and 100 GeV, respectively.

The EFT Lagrangian with complete and independent dim-7 operators describing a pair of the Dirac DM $\chi$ field interacting with the quarks, gluons and photon is given by
\begin{eqnarray}
\mathcal{L}^{\rm dim-7}_\chi=\sum_{i,q} \mathcal{C}_i^7 \mathcal{O}_{i,q}^7 + \mathcal{L}(G^a_{\mu\nu}) + \mathcal{L}(F_{\mu\nu}),
\end{eqnarray}
where the first terms are four-fermion interactions whose Wilson coefficients are assumed to be quark flavor universal, $\mathcal{C}_i^7=\Lambda_i^{-3}$. The last two terms consist of eight operators that couple DM to a pair of gluon or photon field strength tensors. Among the four-fermion operators there are six commonly considered ones, i.e., $m_q\bar\chi\mathcal{O}_\chi\chi \bar{q}\mathcal{O}_q q$ with $\mathcal{O}_{\chi,q}\in\{1,~i\gamma_5,~\sigma^{\mu\nu}\}$, which are suppressed for light quarks. Recently, it was pointed out that there exist additional eight four-fermion operators with a derivative acting on the DM fields as shown in Table~\ref{tab:operator}~\cite{Brod:2017bsw}, where $\bar{\chi}i\overleftrightarrow{\partial_\mu}\chi = \bar{\chi}i\partial_\mu\chi- \bar{\chi}i\overleftarrow{\partial_\mu}\chi$. Since the phenomenological studies of these new operators are still lacking, we will fill this gap.

\begin{table}[h]
\begin{center}
\begin{tabular}{|l|c|}
        \hline
        Operator $\mathcal{O}_{i,q}^7$ & Coefficient $\mathcal{C}_i^7$\\
        \hline
        ${\rm D_{15}}: \bar{\chi}i \overleftrightarrow{\partial_\mu} \chi \bar{q}\gamma^\mu q$ & $1/\Lambda_{15}^3$ \\
        \hline
        ${\rm D_{16}}: \bar{\chi}i\gamma_5 i \overleftrightarrow{\partial_\mu} \chi \bar{q}\gamma^\mu q$ & $1/\Lambda_{16}^3$ \\
        \hline
        ${\rm D_{17}}: \bar{\chi}i \overleftrightarrow{\partial_\mu} \chi \bar{q}\gamma^\mu \gamma_5 q$ & $1/\Lambda_{17}^3$ \\
        \hline
        ${\rm D_{18}}: \bar{\chi}i\gamma_5 i \overleftrightarrow{\partial_\mu} \chi \bar{q}\gamma^\mu \gamma_5 q$ & $1/\Lambda_{18}^3$ \\
        \hline
        ${\rm D_{19}}: \partial_\mu (\bar{\chi}\sigma^{\mu\nu}\chi) \bar{q}\gamma_\nu q$ & $1/\Lambda_{19}^3$ \\
        \hline
        ${\rm D_{20}}: \partial_\mu (\bar{\chi}\sigma^{\mu\nu}i\gamma_5\chi) \bar{q}\gamma_\nu q$ & $1/\Lambda_{20}^3$ \\
        \hline
        ${\rm D_{21}}: \partial_\mu (\bar{\chi}\sigma^{\mu\nu}\chi) \bar{q}\gamma_\nu \gamma_5 q$ & $1/\Lambda_{21}^3$ \\
        \hline
        ${\rm D_{22}}: \partial_\mu (\bar{\chi}\sigma^{\mu\nu}i\gamma_5\chi) \bar{q}\gamma_\nu \gamma_5 q$ & $1/\Lambda_{22}^3$ \\
        \hline
\end{tabular}
\end{center}
\caption{New four-fermion dim-7 operators and associated energy scales.}
\label{tab:operator}
\end{table}

In this work we confront the theoretical predictions of these new dim-7 operators with the experimental observables and constraints to infer the most probable mass of DM and its interaction strengths with ordinary matter. We will also specify distinct features of these operators against other operators in DM searches. This paper is organized as follows. In Sec.~\ref{sec:Cons}, we study the experimental observables and constraints for these operators, and their allowed regions of the cutoff scale and DM mass are given in Sec.~\ref{sec:Res}. Finally, we summarize our main results in Sec.~\ref{sec:Con}.

\section{Observables and Constraints}
\label{sec:Cons}

In this section we discuss the experimental observables and constraints for the new dim-7 operators with a derivative in DM currents, including thermal relic abundance, indirect and direct detection, and LHC bounds.

\subsection{Relic Density}

The thermally averaged cross sections of DM pair annihilation into a quark pair through each new dim-7 operator at a time are, to the leading order in DM velocity $v$,
\begin{eqnarray}
\langle \sigma v\rangle_{\rm D_{15}}&=&
{N_C m_\chi^4\over 24\pi \Lambda_{15}^6}v^4\sum_q \Theta(m_\chi-m_q)(3-\beta_q^2)\beta_q,
\\
\langle \sigma v\rangle_{\rm D_{16}}&=&
{N_C m_\chi^4\over 6\pi \Lambda_{16}^6}v^2\sum_q \Theta(m_\chi-m_q)(3-\beta_q^2)\beta_q\nonumber \\
&=&\langle\sigma v\rangle_{\rm D_{20}}
(\Lambda_{16}\to\Lambda_{20}),
\\
\langle \sigma v\rangle_{\rm D_{17}}&=&
{N_C m_\chi^4\over 12\pi\Lambda_{17}^6}v^4\sum_q \Theta(m_\chi-m_q)\beta_q^3,
\\
\langle \sigma v\rangle_{\rm D_{18}}&=&
{N_C m_\chi^4\over 3\pi \Lambda_{18}^6}v^2\sum_q \Theta(m_\chi-m_q)\beta_q^3\nonumber \\
&=&\langle \sigma v\rangle_{\rm D_{22}}
(\Lambda_{18}\to \Lambda_{22}),
\\
\langle \sigma v\rangle_{\rm D_{19}}&=&
{2N_C m_\chi^4\over \pi \Lambda_{19}^6}\sum_q \Theta(m_\chi-m_q)(3-\beta_q^2)\beta_q,
\\
\langle \sigma v\rangle_{\rm D_{21}}&=&
{4N_C m_\chi^4\over \pi \Lambda_{21}^6}\sum_q \Theta(m_\chi-m_q)\beta_q^3,
\end{eqnarray}
where $\beta_q=\sqrt{1-m_q^2/m_\chi^2}$, $N_C=3$ for quarks, and $\Theta$ is the Heaviside function for only taking into account on-shell two-body annihilations. We assume all kinematically accessible quarks in the final state of the DM annihilation. Note that the annihilation rates for operators D$_{16}$ and D$_{18}$ are equivalent to those of D$_{20}$ and D$_{22}$, respectively, when all of the four fermions are on-shell by making repeated use of the equations of motion. The velocity scaling for these annihilation cross sections is collected in the second column of Table~\ref{tab:sigmav}. One can see that they display various types of velocity dependence in the annihilation rate. The annihilation rates are $d$-wave for operators D$_{15}$ and D$_{17}$, and so are proportional to the fourth power of DM velocity $v^4$. Operators D$_{16}$, D$_{18}$, D$_{20}$ and D$_{22}$ however have a $p$-wave term.

\begin{table}[h]
\begin{center}
\begin{tabular}{|c|c|c|}
        \hline
        Operator $\mathcal{O}_{i,q}^7$ & Anni. $\langle\sigma v\rangle$ & NR operator $\mathcal{O}^{N}_i$
        \\
        \hline
        ${\rm D_{15}}$ & $\mathcal{O}(v^4)$ & $1$
        \\
        \hline
        ${\rm D_{16}}$ & $\mathcal{O}(v^2)$ & $\mathbf{s}_\chi\cdot \mathbf{q}$
        \\
        \hline
        ${\rm D_{17}}$ & $\mathcal{O}(v^4)$ & $\mathbf{s}_N\cdot \mathbf{v_\bot}$
        \\
        \hline
        ${\rm D_{18}}$ & $\mathcal{O}(v^2)$ & $(\mathbf{s}_\chi\cdot\mathbf{q})
        (\mathbf{s}_N\cdot \mathbf{v_\bot})$
        \\
        \hline
        ${\rm D_{19}}$ & $\mathcal{O}(1)$ &
        $\mathbf{q}^2$, $(\mathbf{s}_\chi\cdot \mathbf{q})(\mathbf{s}_N\cdot\mathbf{q})$,
        \\
         &  &
        $\mathbf{s}_\chi\cdot(\mathbf{v_\bot}\times \mathbf{q})$, $\mathbf{q}^2 \mathbf{s}_\chi\cdot \mathbf{s}_N$
        \\
        \hline
        ${\rm D_{20}}$ & $\mathcal{O}(v^2)$ & $\mathbf{s}_\chi\cdot \mathbf{q}$
        \\
        \hline
        ${\rm D_{21}}$ & $\mathcal{O}(1)$ & $\mathbf{s}_\chi\cdot(\mathbf{s}_N\times \mathbf{q})$
        \\
        \hline
        ${\rm D_{22}}$ & $\mathcal{O}(v^2)$ & $(\mathbf{s}_\chi\cdot\mathbf{q})(\mathbf{s}_N\cdot \mathbf{v_\bot})$
        \\
        \hline
\end{tabular}
\end{center}
\caption{Velocity scaling of annihilation cross sections $\langle\sigma v\rangle$ and NR DM-nucleon operators $\mathcal{O}^{N}_i$ for dim-7 operators considered in this work. $\mathbf{s}_\chi$ ($\mathbf{s}_N$) is the DM (target nucleon) spin, and $\mathbf{q}$ and $\mathbf{v_\bot}$ are scattering exchange momentum and velocity defined in Ref.~\cite{Brod:2017bsw}.}
\label{tab:sigmav}
\end{table}

The thermal DM relic abundance is determined by the equation~\cite{Gelmini:1990je}
\begin{eqnarray}
\Omega_\chi h^2 = {1.07\times 10^9 \ {\rm GeV}^{-1}\over M_{\rm Pl}} {x_F\over \sqrt{g_\ast}}{1\over a+3b/x_F+20c/x_F^2},
\label{relic}
\end{eqnarray}
for the expansion of annihilation cross section $\langle \sigma v\rangle\sim a+bv^2+cv^4$. Here, $M_{\rm Pl}\approx 1.22\times 10^{19}$ GeV is the Planck mass, $h$ is the Hubble parameter, $g_\ast$ is the number of relativistic degrees of freedom, and $T_F$ is the freeze-out temperature appearing in $x_F=m_\chi/T_F$. We vary $x_F$ and $g_\ast$ in the range of $20<x_F<30$~\cite{Jungman:1995df,Cao:2009uw} and $80<g_\ast<100$~\cite{Schafer:2003vz}, respectively, and adopt the relic abundance measured by Planck, i.e. $\Omega_\chi h^2 = 0.1199\pm 0.0027$~\cite{Ade:2015xua}. Note that these choices are rather simplistic but are sufficient to estimate the relic density in the context of EFT~\cite{Balazs:2017hxh}.

\subsection{Indirect Detection}

Dwarf galaxies are bright targets to search for DM annihilation through gamma rays. The Fermi Large Area Telescope (LAT) has searched for gamma ray emission from the dwarf spheroidal satellite galaxies (dSphs) of the Milky Way but detected no excess. The Fermi-LAT thus set an upper limit on the DM annihilation cross section from a combined analysis of multiple Milky Way dSphs~\cite{Ackermann:2015zua,Fermi-LAT:2016uux}. For individual dwarf galaxy targets, the Fermi-LAT collaboration tabulated the delta-log-likelihoods as a function of the energy flux bin-by-bin. The gamma ray energy flux from DM annihilation in the $j$th energy bin is given by
\begin{eqnarray}
\Phi^E_{j,k}(m_{\chi},\langle\sigma v\rangle,J_k)
=\frac{\langle\sigma v\rangle}{16\pi m_{\chi}^2}J_k
\int^{E^{\rm max}_j}_{E^{\rm min}_j}E
\frac{dN_\gamma}{dE}dE,
\end{eqnarray}
where $J_k$ is the $J$ factor for the $k$th dwarf and $dN_\gamma/dE$ describes the gamma-ray spectrum from DM annihilation. The energy flux only depends on $m_{\chi}$, $\langle \sigma v\rangle$ and $J_k$, and is calculable for DM annihilation processes given by the above EFT operators. The likelihood for the $k$th dwarf is
\begin{eqnarray}
&&\mathcal{L}_k(m_{\chi},\langle \sigma v\rangle,J_k)=\nonumber \\
&&\mathcal{L}_J(J_k|\bar{J}_k,\sigma_k)\prod_j \mathcal{L}_{j,k}(\Phi^E_{j,k}(m_{\chi},\langle \sigma v\rangle,J_k)),
\end{eqnarray}
where $\mathcal{L}_{j,k}$ is the likelihood tabulated by the Fermi-LAT for each dwarf and calculated gamma-ray flux and the uncertainty of the $J$ factors is taken into account by profiling over $J_k$ in the likelihood below~\cite{Ackermann:2015zua}
\begin{eqnarray}
&&\mathcal{L}_J(J_k|\bar{J}_k,\sigma_k)={1\over \ln(10)J_k\sqrt{2\pi}\sigma_k}\nonumber \\ 
&&\times \exp\bigg[-\frac{1}{2\sigma_k^2}
\Big(\log_{10}(J_k)-\log_{10}(\bar{J}_k)\Big)^2\bigg],
\end{eqnarray}
with the measured $\bar{J}_k$ and error $\sigma_k$. Then one can perform a joint likelihood for all dwarfs
\begin{eqnarray}
\mathcal{L}(m_{\chi},\langle \sigma v\rangle,\mathbb{J})
=\prod_k\mathcal{L}_k(m_{\chi},\langle\sigma v\rangle,J_k),
\end{eqnarray}
where $\mathbb{J}$ is the set of $J_k$ factors. In our implementation we adopt the corresponding values of $\mathcal{L}_{j,k}$ and $\bar{J}_k, \sigma_k$ for 19 dwarf galaxies considered in Ref.~\cite{Fermi-LAT:2016uux}.

As Fermi-LAT found no gamma ray excess from the dSphs, for a given $m_{\chi}$, one can set an upper limit on the DM annihilation cross section by taking $J$ factors as nuisance parameters in the maximum likelihood analysis. We follow Fermi's approach and take the delta-log-likelihood as below
\begin{eqnarray}
-2\Delta \ln \mathcal{L}(m_{\chi},\langle \sigma v\rangle)=-2\ln\left({\mathcal{L}(m_{\chi},\langle \sigma v\rangle,\widehat{\widehat{\mathbb{J}}})\over \mathcal{L}(m_{\chi},\widehat{\langle \sigma v\rangle},\widehat{\mathbb{J}})}\right),
\end{eqnarray}
where $\widehat{\langle \sigma v\rangle}$ and $\widehat{\mathbb{J}}$ maximize the likelihood while $\widehat{\widehat{\mathbb{J}}}$ maximizes the likelihood for given $m_{\chi}$ and $\langle \sigma v\rangle$. The 95\% C.L. upper limit on the annihilation cross section for a given $m_{\chi}$ is determined by demanding $-2\Delta \ln\mathcal{L}(m_{\chi},\langle \sigma v\rangle)\leq 2.71$. We obtain the spectrum of photons induced by annihilation into quarks using the PPPC4DMID code~\cite{Cirelli:2010xx} and perform the likelihood analysis using Minuit~\cite{James:1975dr}. Once the annihilation cross section obtained from a certain set of $m_\chi$ and $\Lambda$ is larger than the limit, we claim the corresponding parameter values are excluded by the Fermi-LAT dSphs measurement.
Due to the suppression by the extremely non-relativistic DM velocity, we expect ID constraints to be relatively weaker for d- and p-wave operators.

\subsection{Direct Detection}

\begin{figure}[h!]
\begin{center}
\includegraphics[scale=1,width=5cm]{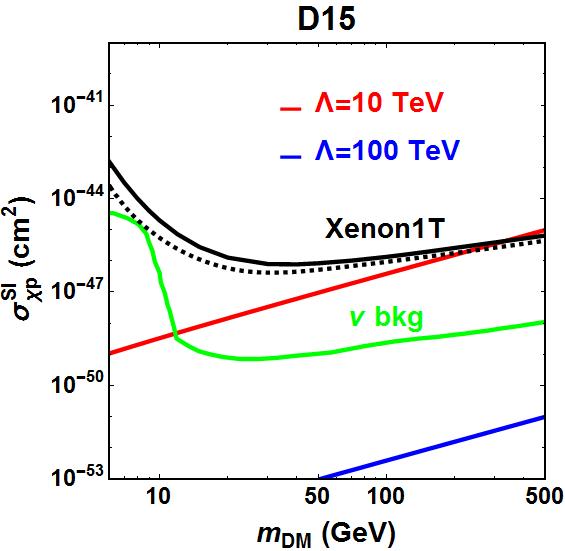}
\end{center}
\caption{SI DM-proton scattering cross section versus DM mass for operator $\rm D_{15}$ at $\Lambda_{15}=10,~100~\textrm{TeV}$ is compared with Xenon1T limits at 90\% CL (solid black~\cite{Aprile:2017iyp} and dashed black~\cite{Aprile:2018dbl}) and neutrino background (green).}
\label{SID15}
\end{figure}

We show in the third column of Table~\ref{tab:sigmav} the non-relativistic (NR) operators of DM scattering off the nucleon induced from the considered dim-7 operators at quark level~\cite{Brod:2017bsw}. One can see that, for operators D$_{16}$-D$_{22}$, the scattering rates are either suppressed by the spin of the target nucleus $\mathbf{s}_N$ or the scattering momentum exchange $\mathbf{q}$ or both, rendering weak DD constraints. Only operator D$_{15}$ leads to non-momentum-suppressed spin-independent (SI) DM-nucleon scattering, and is thus highly constrained by the direct DM detection. The NR reduction of the operator D$_{15}$ to the DM-nucleon level  is~\cite{DeSimone:2016fbz,Brod:2017bsw}
\begin{eqnarray}
&&\mathcal{C}_{\rm D_{15}} \mathcal{O}_{\rm D_{15}} \to \mathcal{C}_{\rm D_{15}}^N \mathcal{O}_1^{N} = 2m_\chi \left(2\mathcal{C}_{\rm D_{15}}+\mathcal{C}_{\rm D_{15}} \right) \mathcal{O}_1^{N}= {6m_\chi \over \Lambda_{15}^3}\mathcal{O}_1^{N}, \nonumber \\
&&{\rm with} \ \mathcal{O}_1^{N}=1_\chi 1_N,
\end{eqnarray}
for the interaction of DM with the nucleon. The SI DM-nucleon scattering cross section is thus given by
\begin{eqnarray}
&&\sigma^{\rm SI}_{\chi N}({\rm D_{15}})={\mu_{\chi N}^2\over \pi}\left(\mathcal{C}_{\rm D_{15}}^N\right)^2={\mu_{\chi N}^2\over \pi}\left({6m_\chi\over \Lambda_{15}^3}\right)^2 \nonumber \\
&&= 4.47\times 10^{-43} \ {\rm cm}^2 \left({\mu_{\chi N}\over 1 \ {\rm GeV}}\right)^2 \left({m_\chi\over 10 \ {\rm GeV}}\right)^2 \left({1 \ {\rm TeV}\over \Lambda_{15}}\right)^6,
\end{eqnarray}
where $\mu_{\chi N}=m_\chi m_N/(m_\chi + m_N)$ is the reduced mass with $m_N$ being the nucleon mass. This
prediction can then be compared directly to the limits set by DD experiments to yield a lower bound on the cutoff scale $\Lambda_{15}$ for a given $m_\chi$. In Fig.~\ref{SID15}, we show the SI DM-proton scattering cross section versus DM mass for different values of $\Lambda_{15}$. For instance, for $\Lambda_{15}=1$ TeV the whole range of $m_\chi > 10$ GeV is excluded by Xenon 1T~\cite{Aprile:2017iyp,Aprile:2018dbl}, while DM with $m_\chi < 300$ GeV can evade the DD limit when $\Lambda_{15}=10$ TeV.

\subsection{LHC Constraints}

The LHC constraints on DM EFT stem from searches for large missing energy events produced alongside with a visible object such as a jet, lepton, or photon, i.e. the so-called mono-X searches. The most stringent constraint for the operators we consider comes from the mono-jet search corresponding to an integrated luminosity of 36.1 fb$^{-1}$ at a centre-of-mass energy of 13 TeV~\cite{Aaboud:2017phn}. In order to estimate the mono-jet constraint on our EFT setups, we create UFO model files using FeynRules~\cite{Alloul:2013bka} and interface them with MadGraph5\_aMC@NLO~\cite{Alwall:2014hca} to generate signal events composed of DM pairs with a jet from initial-state radiation. The signal events are then passed to Pythia~\cite{Sjostrand:2006za} and Delphes~\cite{deFavereau:2013fsa} for parton shower and detector simulation, respectively. Following the event selection in Ref.~\cite{Aaboud:2017phn}, we require the leading jet satisfying $p_T>250$ GeV and $|\eta|<2.4$ and the missing transverse momentum with $E_T^{\rm miss}>250$ GeV. Ref.~\cite{Aaboud:2017phn} provides the observed 95\% confidence level (CL) upper limit on the visible cross section, defined as the product of cross section and efficiency corresponding to the above selection cuts. Once the visible cross section obtained from a certain value of $m_\chi$ and $\Lambda$ is larger than the limit, we claim the corresponding parameters are excluded by the mono-jet search at 95\% CL.

\begin{figure}[h!]
\begin{center}
\includegraphics[scale=1,width=6cm]{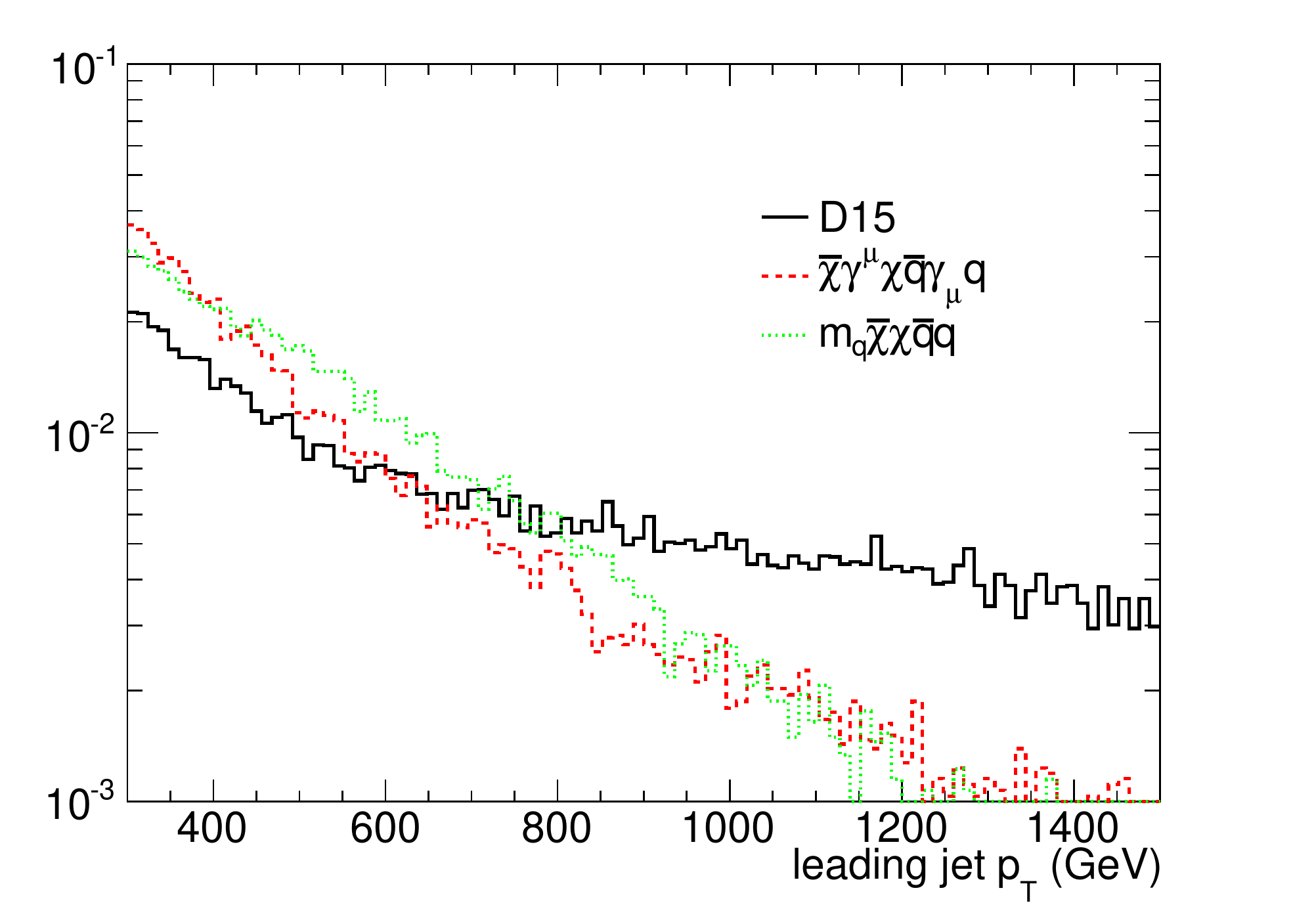}
\end{center}
\caption{Normalized leading jet $p_T$ for D$_{15}$ operator (black solid), dimension-6 operator $\bar{\chi}\gamma^\mu\chi\bar{q}\gamma_\mu q$ (red dashed) and dimension-7 operator $m_q\bar{\chi}\chi\bar{q}q$ (green dotted), assuming $m_\chi = 100$ GeV.
}
\label{pt}
\end{figure}

In Fig.~\ref{pt} we compare the normalized distributions in the transverse momentum of the leading jet for the D$_{15}$ operator (black), a dim-6 operator $\bar{\chi}\gamma^\mu\chi\bar{q}\gamma_\mu q$ (red), and a dim-7 operator $m_q\bar{\chi}\chi\bar{q}q$ (green), assuming $m_\chi = 100$ GeV at 13 TeV LHC. The signal distribution of the new dim-7 operator with a derivative does not decrease as fast as the other two in the high energy region due to the derivative enhancement~\cite{Cai:2018cog}.

\section{Results}
\label{sec:Res}

\begin{figure}[h!]
\begin{center}
\minigraph{4.2cm}{-0.05in}{(a)}{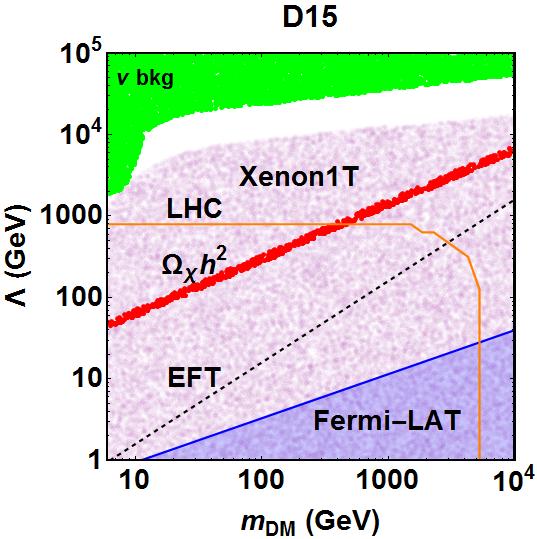}
\minigraph{4.2cm}{-0.05in}{(b)}{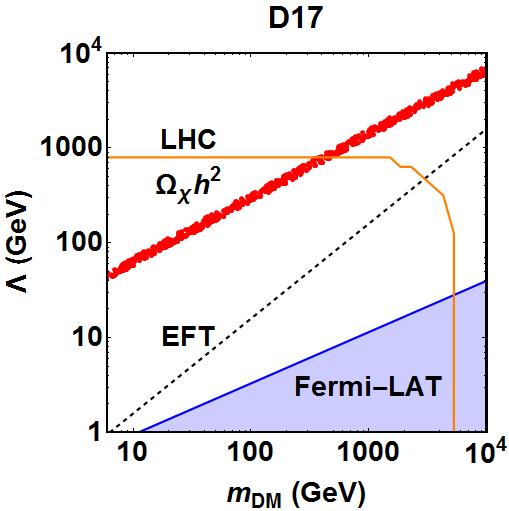}
\end{center}
\caption{The allowed region of cut-off scale $\Lambda$ vs. $m_\chi$ by Planck (red band) for operators $\rm D_{15}$ (a) and $\rm D_{17}$ (b). EFT is valid above the dashed line. The blue region is excluded by the null measurement of dwarf galaxies by Fermi-LAT. The excluded region by Xenon1T (purple) and the region below the neutrino background (green) are also shown for operator $\rm D_{15}$. The orange curve represents the LHC bound.}
\label{v4}
\end{figure}

\begin{figure}[h!]
\begin{center}
\minigraph{4.2cm}{-0.05in}{(a)}{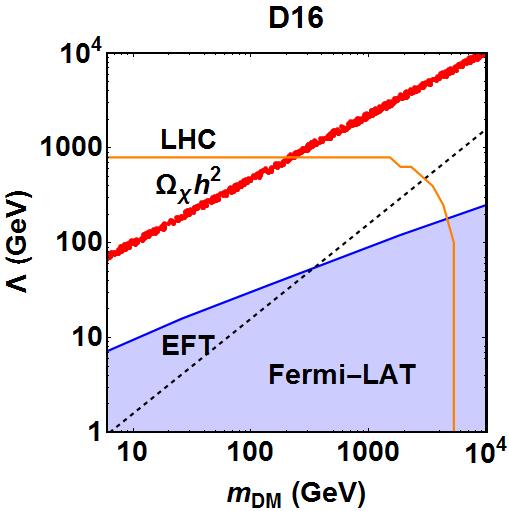}
\minigraph{4.2cm}{-0.05in}{(b)}{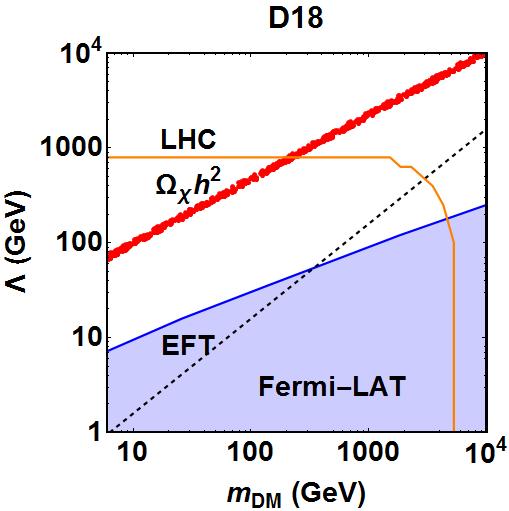}
\end{center}
\caption{Results for operators ${\rm D}_{16} ({\rm D}_{20}$) and ${\rm D}_{18} ({\rm D}_{22}$), as labeled in Fig.~\ref{v4}.
}
\label{v2}
\end{figure}

\begin{figure}[h!]
\begin{center}
\minigraph{4.2cm}{-0.05in}{(a)}{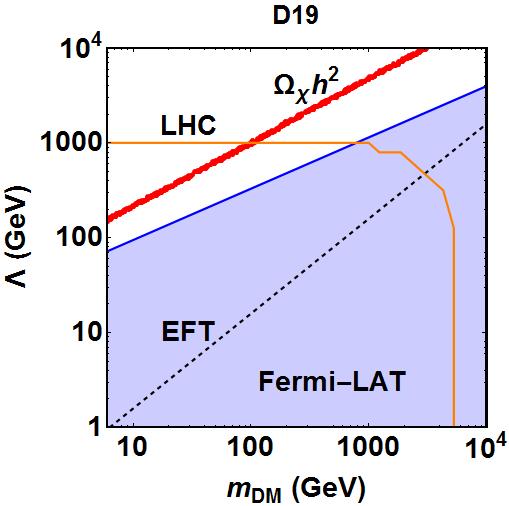}
\minigraph{4.2cm}{-0.05in}{(b)}{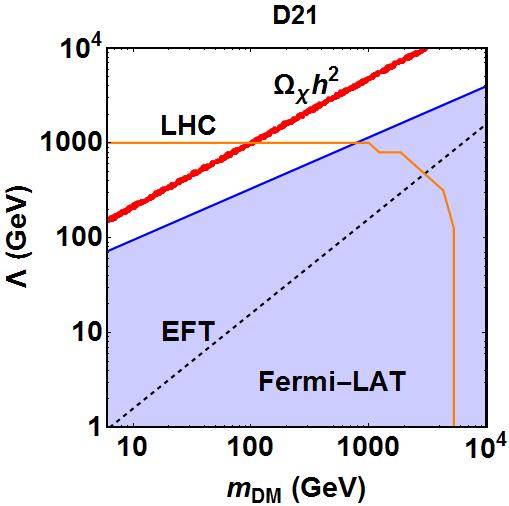}
\end{center}
\caption{Results for operators ${\rm D}_{19}$ (a) and ${\rm D}_{21}$ (b), as labeled in Fig.~\ref{v4}.}
\label{v0}
\end{figure}

In this section, for each operator in Table~\ref{tab:operator} we show its allowed region in the cut-off scale $\Lambda_i$ versus $m_\chi$ by the observables discussed in the previous section. Figs.~\ref{v4}, \ref{v2}, and \ref{v0} correspond to the operators resulting in $d$-, $p$- and $s$-wave annihilation rates, respectively.

The correct thermal DM relic abundance with corresponding $\Lambda$ and $m_\chi$ is given by the red band. The band is derived from Eq.~(\ref{relic}) and reflects the assumed ranges of values for $x_F$ and $g_\ast$. As expected before, the more the annihilation rate is suppressed, the weaker the ID constraint such as the exclusion by Fermi-LAT dSphs becomes. Most severely, as indicated by the blue squares, Fermi-LAT excludes a majority of space below $\Lambda_i\simeq 7$ TeV for operators D$_{19}$ and D$_{21}$.

The mono-jet search at 13 TeV LHC excludes the parameter space to the left of the orange solid line that essentially amounts to a lower limit on the EFT scale $\Lambda_i\lesssim 1$ TeV. This LHC constraint is more sensitive to the low $m_\chi$ region, thus complementary to the indirect detection. Besides, the limit of scattering cross section from Xenon 1T at 90\% CL severely constrains the operator D$_{15}$ such that only the blank band in the left panel of Fig.~\ref{v4} remains to be explored by future direct detection experiments.

Finally, the EFT approximation is valid above the black dotted lines, i.e. roughly for $\Lambda_i>m_\chi/(2\pi)$. The region yielding a correct relic density is compatible with EFT validity. Larger couplings will violate perturbative unitarity whence the EFT expansion breaks down and cannot give a reliable description of an underlying theory.

In summary, to avoid overproduction of DM and ensure the validity of the EFT approximation, the viable $\Lambda_i$--$m_\chi$ region must fall between the red band and the dotted line. Furthermore, the energy scale $\Lambda_i$ has to be greater than about 1 TeV to satisfy the LHC bound. In particular, this squeezed region is entirely excluded by direct detection for D$_{15}$ and mostly excluded by indirect detection for D$_{19}$ and D$_{21}$.

\section{Conclusion}
\label{sec:Con}

In this work we have investigated new dimension-7 operators in effective field theory for Dirac fermionic dark matter. These operators involve a derivative in the DM currents and their phenomenology has not yet been studied in the literature. We discussed the experimental observables and constraints for these operators to confine the DM mass and relevant energy scale. We found that these operators induce various $s$-, $p$- and $d$-wave annihilation rates and are thus, to different extents, constrained by indirect DM detection such as the Fermi-LAT dSphs. In spite of this, the correct thermal relic abundance can be achieved in the parameter space allowed by indirect detection. The mono-jet search at 13 TeV LHC excludes the parameter space with energy scale $\Lambda_i\lesssim 1$ TeV and $m_\chi\lesssim 8$ TeV. And only one of the operators gives non-momentum-suppressed spin-independent DM-nucleon scattering, and is thus highly constrained by direct detection experiments.

\section*{Acknowledgments}

TL thanks Thomas Jacques for helpful discussions. This work was supported in part by the Grants No. NSFC-11575089 and No. NSFC-11025525, by The National Key Research and Development Program of China under Grant No. 2017YFA0402200, and by the CAS Center for Excellence in Particle Physics (CCEPP).
T.L. is supported by ``the Fundamental Research Funds for the Central Universities'', Nankai University (Grant Number 63191522, 63196013).


\end{document}